\documentclass[useAMS,usenatbib,psfig]{mn2e}
\usepackage{graphicx}

\title[3D SPH Simulations of Shocks in Accretion Flows around black
holes] {3D SPH Simulations of Shocks in Accretion Flows around
black holes}
\author[G. Gerardi, D. Molteni, V. Teresi]{G. Gerardi$^1$ \thanks{E-mail:
gerardi@difter.unipa.it (GG)}, D. Molteni$^1$ and V. Teresi$^{1}$ \\
$^{1}$Dipartimento di Fisica e Tecnologie Relative, Universit$\grave{a}$
di Palermo, Viale delle Scienze, Palermo, 90128, Italy\\
}
\begin{document}



\maketitle

\label{firstpage}

\begin{abstract}
We present the simulation of 3D time dependent flow of rotating
ideal gas falling into a Schwarzschild black hole. It is shown
that also in the 3D case steady shocks are formed in a wide range
of  parameters (initial angular momentum and thermal energy). We
therefore highlight the stability of the phenomenon of shock
formation in sub keplerian flows onto black holes, and reenforce
the role of the shocks in the high luminosity emission from black
hole candidates. The simulations have been performed using a
parallelized code based on the Smoothed Particles Hydrodynamics
method (SPH). We also discuss some properties of the shock problem
that allow its use as a quantitative test of the accuracy of the
used numerical method. This shows that the accuracy of SPH is
acceptable although not excellent.
\end{abstract}


\section{Introduction}
As Bondi showed in his historical paper \citep{Bondi52} the
occurrence of shocks in accretion phenomena on compact stellar
objects is quite common. Since the radial speed can become
supersonic the halting of the flow at a rigid wall leads
necessarily to shock formation.

Nevertheless the possibility of radial shock formation around
black holes is less obvious since this compact object has no rigid
wall. It is true that pre-heating of the infalling gas by
radiation from below or other non adiabatic processes can
influence the properties of the falling gas and allow shock
solutions even in the spherically symmetric regime
\citep{Babul89}.

However, if the gas is falling with some rotation, the centrifugal
force can act as a "wall" and therefore trigger the shock
formation even in the simple case of ideal gas. This fact was
indeed suggested by the excellent pioneer general relativistic
simulations of Hawley, Smarr and Wilson \citep{Hawley84}. The
question of the stability of the shock is critically related to
the initial angular momentum of the gas, since for large values a
standard keplerian disk will be formed, while for low values
favorable conditions for standing shocks will occur.

That such shocks can effectively stay at a fixed position has been
put forward by S. Chakrabarti \citep{Chakrabarti90}, see also
\citet{Chakrabarti98}. It has been successively confirmed by the
analytical work of Nakayama \citep{Nakayama94} and by numerical
simulation experiments by Chakrabarti and Molteni
\citep{Chakrabarti93} and following papers. The stability of the
shock location is very relevant since, if it is stable, then the
shock "engine" is always acting: it is not a transient phenomenon.
Further studies, based on numerical simulations in 2D, axis
symmetric configurations, revealed different kinds of oscillating
behavior over the basic stable regime, see the results in
\citet{Molteni96}, \citet{Molteni96b}, \citet{Molteni99}.

Recently the shock scenario has been shown to be very promising
also to explain quasi periodic oscillation processes (QPO)
occurring in BH candidates \citep{Molteni01}.

Since all previous studies have been performed in 2D
configurations (axial symmetric or in the Z=0 plane), it is
natural to investigate the stability of the shocks in full 3D
cases. Furthermore we add that this physical problem can be also
considered a very critical test in the field of computational
fluid dynamics.

The work proceeds as follows: in section 2 we resume the basic
physics leading to shock formation and derive a simple formula to
obtain the physical parameters necessary to generate the shocks,
in section 3 we revise the numerical method and show how a test of
any numerical code can be set up and in section 4 we show and
discuss the new 3D simulation results.

\section{Shock formation in unviscous sub-keplerian flows}

Let us resume the very basic physical ingredients of the
phenomenon. We suppose that an unviscous gas is falling from very
large distance onto a black hole. We will adopt the Paczy\'nski \&
Wiita potential \citep{Pac80}  to mimic the Schwarzschild Black
Hole force. It is given by :
\begin{equation}
\Psi \left( r\right) =-\frac{GM_{\star }}{r-r_g}
\end{equation}
where $ r_g= \frac{2 \cdot G \cdot M}{c^{2}} $  is the
Schwarzschild radius.
The basic physical effect can be easily
understood in terms of "classical" physics and it is well known
that the Paczy\'nski \& Wiita force reproduces many relativistic
effects with high accuracy.

To obtain the steady state solution we might integrate the
differential equations for mass, momentum and energy
conservations. In this case one has to start the space integration
from the sonic point as explained in the work by Chakrabarti
\citep{Chakrabarti90}. However, in the case of unviscous flow, it
is easy to find an algebraic implicit solution, since the total
energy is conserved (the Bernoulli theorem) and it can be fairly
exploited to close the system of equations and find the solutions.

\subsection{1D Case}
Let us treat first the 1D case. This derivation already appeared
in the appendix of  \citet{Molteni99}. We repeat here the
derivation pointing also to a  peculiar aspect of the solution
especially relevant to test code accuracy. By 1D we mean that we
are in an axially symmetric situation and the disk has no vertical
extension ($v_z=0$, $Z=0$). Note that the solution is obviously
valid {\it{also}}  in 2D dimensions with $Z=0$ (i.e. in the plane
XY) if the symmetric condition is maintained. Both in the case of
1D simulations in cylindrical coordinates and in 2D simulations
(but with  Z=0) the difference between the theoretical shock
position and the one resulting from simulations depends only on
the numerical quality of the code; indeed there is no physical
cause intervening to modify the results. Instead, in the 2D or 3D
cases (with real vertical extension of the flow), the theoretical
procedure to derive the shock position needs the assumption of
vertical equilibrium, that, as we will see, is not always
satisfied.

 We assume also that the accreting gas is ideal (no viscosity and no
cooling process is occurring). The angular momentum per unit mass
of the flow, $\lambda$, must be conserved. In steady state regime
we have also conservation of mass given by the equation:

\begin{equation}
\label{1} \dot{M}=-r\rho v_r=const
\end{equation}

Here $\rho$ is the surface density. In a conservative body force
field with a potential $\Psi \left(r\right)$, also the pressure
plays the role of a 'potential' and we have

\begin{equation}\label{Bernoulli}
  \frac{1}{2} v\left( r\right)^{2}+\epsilon\left( r\right) +\frac{P\left( r\right) }{\rho \left( r\right) }+\Psi
\left( r\right)= \frac{1}{2}v\left( r\right) ^2+\frac{a\left( r\right) ^2}{\left( \gamma -1\right) }+\Psi \left(
r\right) = B
\end{equation}

Here $\gamma$ is the  adiabatic gas constant,  $B$ is the
Bernoulli constant (the thermal energy at infinity), $a$ is the
sound speed $a=\sqrt{\frac{\gamma\cdot P}{\rho}},$ $\epsilon$ is
the internal energy per unit mass, $v$ is the speed of the flow
and $\lambda $ is the angular momentum, so that $\frac{1}{2}v^2=
\frac 1 2 v_\phi ^2+\frac 1 2v_r^2= \frac{\lambda
^2}{2r^2}+\frac{1}{2}v_r^2$.

If the entropy is conserved (no viscosity, no cooling, the
presence of the shock will be taken into account few lines below)
the polytropic relations are valid, $P/ \rho^{\gamma}=P_{0}/
\rho_{0}^{\gamma}$, so the relation between the density and sound
speed $ \rho= K\cdot a^{2/(\gamma-1)}$, where $ K= \rho _0/a
_0^{2/(\gamma-1)}$, is also valid.

We have three unknown quantities $\rho$, $v_r$, $a$, and three
equations. The solutions are functions of $r$ with angular
momentum and Bernoulli constant as parameters determining the
specific shape of the solution.

Using the radial Mach number  $m=-\frac{v_r}a$, resolving  $a$
from the Bernoulli relation, and putting all terms in the
continuity equation, we have the following implicit solution for
the Mach number:
$$\dot{M}=r\cdot m\cdot a \cdot K \cdot a ^{\frac{2}{\gamma-1}}
  \propto f(m)\cdot A\left( r,B,\lambda \right)$$
where $f$ is function  only of the mach number:%
$$
f\left( m\right) =\frac m{\left[ \frac{m^2}2+\frac
1{\gamma-1}\right]^{ \frac{\gamma +1}{2\left( \gamma -1\right) }}}
$$
and $A$ is function only of $r$ (B and $\lambda$ are parameters) :%
$$
A\left( r,B,\lambda \right) = r \left[ B- V\left( r,\lambda\right)
\right]^{ \frac{\gamma +1}{2\left( \gamma -1\right) }}
$$
where
$$ V\left( r,\lambda \right) =\frac{\lambda ^2}{2r^2}-\frac{GM_{\star }}{r-r_g}$$
is the effective body force potential (gravitational plus
centrifugal).

For a given $ \dot{M} $ this equation can be easily solved. The
$f(m)$ function has a single maximum at $m=1$, and it can be
inverted both in the subsonic $0\le m < 1$ and supersonic $ m > 1
$ regions. The $A$ function has in general two local minima at
$r_1$ and $r_2$ with $r_1 < r_2$, which can be determined by
solving the algebraic equation $dA/dr=0$ numerically (high
$\lambda$ values increase the centrifugal barrier and produce
minima close to the BH).

At large distances from the BH, the Mach number is $m\ll 1$, while
close to the horizon it is $m\gg 1$. Since the $f(m)\cdot A\left( r,B,\lambda \right)$
product must be constant along the flow, the maximum of $f(m)$
must be at one of the minima of $A(r)$, i.e. at $r_1$ or $r_2$.
Therefore we can have two isentropic solutions $m_1(r)$ and
$m_2(r)$
\begin{equation}\label{q-mach}
 m_{1,2}(r)=f^{-1}\left[\frac{f(1)A_{\lambda,B}(r_{1,2})}{A_{\lambda,B}(r)}\right]
\end{equation}

A standing shock can occur in the solution at $r_{shock}$ if the
values $m_2(r_{shock})>1$ and $m_1(r_{shock})<1$ are related by
the Hugoniot relation
\begin{equation}\label{q-machshock}
 m_1(r_{shock})=h(m_2(r_{shock}))
 \end{equation}

where the Hugoniot relation for an hypothetical shock  at a
generic radius is given by

\begin{equation}\label{q-machshock}
h(m_2(r))=\left[\frac{2+(\gamma-1)m_2(r)^2}{2\gamma
m_2(r)^2-(\gamma-1)}\right]^{1/2}
\end{equation}

In general there can be two possible shock positions, but only the
outer one is stable as shown by the numerical simulations of
Chakrabarti and Molteni \citep{Chakrabarti93} and by the
analytical work of Nakayama \citep{Nakayama94}.

Figure 1 clarifies the procedure for a generic case. The
supersonic line corresponds to the $m_1(r)$ solution while
$m_2(r)$ corresponds to the subsonic branch.

\begin{figure}
\begin{center}
\includegraphics[scale = 0.30,angle = 270.0]{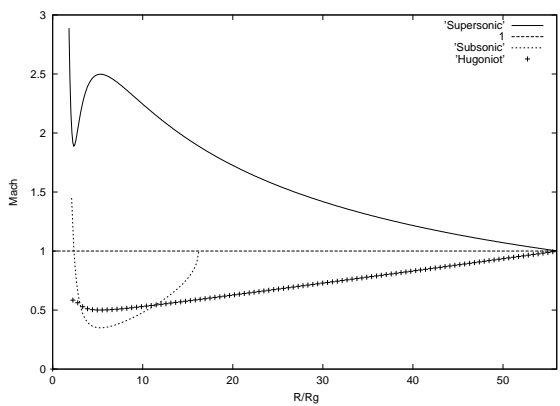}
\label {Ansol} \caption{Analytical solution}
\end{center}
\end{figure}

\subsection{ Fictious and true rotation}
 Let us note that the solution depends on the A function that contains the term:

\begin{equation}
 B- \frac{\lambda ^2}{2r^2}+ \frac{ GM_{\star }} {r-r_g}
\end{equation}

So, if we give to the flow an angular momentum lower than the
theoretical one, but we add an equivalent centrifugal force, the
resulting solution does not change. For example, we can obtain the
same shock location if we give {\it no real  rotation} to the flow
and add to the gravitational force a fictitious force derived from
the centrifugal potential $\frac{\lambda ^2}{2r^2}$, or viceversa,
we can inject matter with an extremely fast real rotation, (even
super keplerian!) and, at the same time, add a centripetal force.

 In this way, playing with real rotation and fictious centrifugal
 forces we can obtain a variety of solutions with the {\it same}
 radial shock position, {\it same} radial Mach number, {\it same}
 density profile. The important point is that the algebraic sum of real
 and fictious forces must be equal to the force required by the theory.
It means that one of the following equations must be satisfied,
according to the case 

\[
\left\{ {{\begin{array}{*{20}c}
 {\lambda_{theor}^{ 2} = \lambda_{real}^{ 2}+ \lambda_{fict}^{2}} \hfill & {\mbox{if}}\hfill & {\lambda_{real}<\lambda_{theor}} \hfill \\
 {\lambda_{theor}^{ 2} = \lambda_{real}^{ 2}- \lambda_{fict}^{2}} \hfill & {\mbox{if}}\hfill & {\lambda_{real}>\lambda_{theor}} \hfill \\
\end{array} }} \right.
\]

where $\lambda_{theor}$ is the angular momentum value required by
the theory to produce a shock at some specified radius. For
consistency, if we define the real rotation as a fractio $f$ of
the theoretical one, i.e. $\lambda_{real}=f\cdot \lambda_{theor}$,
then the fictious force to be added is given by:
\[
\left\{ {{\begin{array}{*{20}c}
 {\lambda_{fict} = \lambda_{theor}\cdot\sqrt{1-f^2}} \hfill & {\mbox{if}}\hfill & {f<1} \hfill \\
 {\lambda_{fict} = \lambda_{theor}\cdot\sqrt{f^2-1}} \hfill & {\mbox{if}}\hfill & {f>1} \hfill \\
\end{array} }} \right.
\]

It is clear that in the fast rotational cases the real shear
motion can modify the results due to the numerical viscosity and
other numerical  code properties:  accuracy, conservative
structure, etc...

We suggest the possibility to quantify the quality, $Q$, of a code
as the ratio of the true rotation angular momentum to the
theoretical one for which the shock is stable and close to the
theoretical position:
$$ Q=1+\frac{\lambda_{real} }{\lambda_{theor}}$$

The obvious minimum requirement is that a code has Q=1 so that,
for no rotation, it produces the shock at the correct position.

\subsection{2D axis symmetric case}

For the 2D case (with real Z extension of the disk) the procedure
is similar. We maintain the hypothesis of axial symmetry. We
assume that the gas falls down in a condition of vertical
equilibrium. Using the same variable names, the mass conservation
equation is given by:
\begin{equation}
\label{1} \dot{M}=- 4 \pi r H \rho v_r=const
\end{equation}
where H is the half thickness of the disk.

The vertical equilibrium hypothesis gives the well known
expression for the half disk thickness:
\begin{equation}
\label{1} H= {\frac {\sqrt {G{  M_{\star }}\,r} \left( r-{  r_g}
\right) {  a}}{ G{  M_{\star }}}}
\end{equation}

The energy equation is still given by:
\begin{eqnarray}
 \nonumber 
 \frac{\lambda
^2}{2r^2}+\frac{1}{2}v_r^2+\epsilon\left( r\right) +\frac{P\left(
r\right) }{\rho \left( r\right) }+\Psi \left( r\right)  &=B
\end{eqnarray}

So again we have three unknown quantities $\rho$, $v_r$, $a$, and
three equations and we may find the solution exactly in the same
way of the 1D case. The difference will be in the exact values of
the Mach function and of the geometrical function, $A(r,
\lambda,B$), but the shape are very similar: two minima in the
geometrical function and one maximum in the Mach function.

The Mach function is now given by:
$$
f\left( m\right) =\frac m {\left[ \frac{m^2}2+\frac 1{\gamma
-1}\right] ^{ \frac{\gamma  }{ \left( \gamma -1\right) }}}
$$
and the new $A$ is function only of $r$ (B and $\lambda$ are parameters) :%
$$
A\left( r,B,\lambda \right) =r^{\frac  3 2} \left( r -r_g \right)
\left[ B- V\left( r,\lambda \right) \right]^ {
\frac{\gamma}{\left( \gamma -1\right) }}
$$
where as previously stated
$$ V\left( r,\lambda \right) =\frac{\lambda ^2}{2r^2}-\frac{GM_{\star }}{r-r_g}$$
is the effective body force potential (gravitational plus
centrifugal).

\section{The physical model and the numerical simulations}
We performed 2D and 3D time dependent simulations of the motion of
an ideal gas around a Schwarzschild black hole. The equations we
integrate are in the lagrangian formulation ($\frac{d}{dt}$ is the
comoving derivative):\\
\newpage
the continuity equation
\begin{equation}\label{continuity}
  \frac{d \rho}{dt} = -\rho\nabla\cdot \mathbf{v}
\end{equation}

the momentum equation
\begin{equation}\label{momentum}
  \frac{d \mathbf{v}}{d t}=-\frac{1}{\rho } \nabla p
\end{equation}

the energy equation
\begin{equation}\label{energy}
  \frac{d \epsilon}{d t} = - \frac{p}{\rho} \nabla\cdot \mathbf{v}
\end{equation}

the  equation of state
\begin{equation}\label{state}
  p= (\gamma -1)\rho \epsilon
\end{equation}
Applying the SPH formalism (see \citet{Monaghan92}), these are
transformed into a set of ordinary differential equations (ODEs).
These ODEs are subsequently integrated numerically using standard
methods.
 We use the usual summation form for the continuity equation:
\begin{equation}\label{cont2}
  \rho_{i}  = \sum_{k=1}^{N} m_{k}W_{ik}
\end{equation}
the momentum equation is given by:
\begin{equation}\label{mom}
   \frac{d\mathbf{v}_{i}}{dt}= -\sum_{ k=1}^{N} m_{k}(\frac{p_{k}}{\rho_{k}^{2}}+
    \frac{p_{i}}{\rho_{i}^{2}}+\Pi_{ik})\cdot\nabla_{i}W_{ik}
\end{equation}
the energy equation is:
\begin{equation}\label{ener}
 \frac{d\epsilon_{i}}{dt}= -\sum_{ k=1}^{N} m_{k}(\frac{p_{k}}{\rho_{k}^{2}}+
  \frac{p_{i}}{\rho_{i}^{2}}+\Pi_{ik})\mathbf{v}_{ik}\cdot \nabla_{i}W_{ik}
\end{equation}
Where $\Pi_{ik}$ is the standard form of the artificial viscosity.

\subsection{Simulations parameters}
Here we show the results of 3D time dependent simulations and add
also the results of 2D (Z=0) simulations to clarify the item of
the code accuracy testing. We focus our attention to a limited set
of parameters which predicts a shock position close to the minimum
radius allowed (around $5 R_g$) \citep{Das2001}. Therefore we are
presenting the results concerning two critical cases. However we
verified that also for other parameter values the shocks are
formed and are stable, in good agreement with the predicted
theoretical position. The physical parameters of the cases we
discuss here are given in Table 1 and they are well inside the
stability region analyzed by \citet{Das2001}.


The variables appearing in the Table 1 have the following meaning:
$ri$ is the inflow radius, v$_{ri }$ and a$_{ri}$ are the radial
speed and sound speed at inflow radius, $B$ is the Bernoulli
constant, $\lambda$ is the angular momentum per unit mass, H is
the vertical disk thickness at the inflow radius, R$_{shock}$ is
the predicted shock position, $f$ is the fractio of the real
rotation of the gas.

The spatial domain (in cylindrical coordinates) is given by
$r=0\div r_i $, $\varphi=0 \div 2\pi $ , $z=-2 r_i \div +2 r_i$.

The results have been obtained with a constant value of the
spatial resolution $h$. This choice gives better results than
those ones obtained with variable $h$; we think that the space
varying $h$ option has a larger amount of numerical viscosity due
to the strong jump of $h$ at the shock location. For 3D runs we used
$h=0.5$ and $\gamma = 4/3$.\\
The gas particles are injected (with the physical properties given
in Table 1) from a set of injection points equally spaced onto the
surface of a cylinder of radius $r_i$ and height H, which is
calculated from the vertical equilibrium assumption. A new
particle is injected every time a sphere of radius $h/2$ around
each injector is found to be empty.

\subsection{Code parallelization}
Since our problem doesn't contain self gravitation, we used a
spatial domain decomposition to parallelize the SPH code. We show
here that this approach is very convenient for the kind of problem
we are studying. Our parallel system consists of a biprocessor
workstations cluster.  As our processors pool has local memory, we
use the well known standard: "\emph{Message Passing Interface}"
(MPI) to exchange data. Our physical problem (accretion disks
around central star) has a space domain with some degree of
central symmetry, so it seems logical to divide the overall domain
into sub domains having this type of symmetry. The parallelization
paradigm we use is the Multiple Instruction Multiple Data one
(MIMD). With this decomposition and paradigms, the code performing
the calculations for one sub domain is the standard serial code
used for one single domain, with minor modifications. Information
exchanges occur always between two consecutive domains.

The three dimensional computational domain is decomposed into
concentrically cylindrical coronae (sub domains) with centers
coincident with the coordinate origin. Figure 2 shows an
horizontal cut view, in a case in which the domain is decomposed
into four sub domains. All the SPH particles, having vector radius
projection (in XY plane) lying between two consecutive cylinders,
are assigned to the same sub domain. The number of sub domains is
equal or multiple of the number of processors in the cluster.

As the SPH particles have a short-range interaction, only the
particles lying within a distance $\leq 2h$ from the limiting
cylinders interact with their analogue particles laying in the
previous cylinder or in the next one.

In any moment each processor manages $n_i$ internal particles plus
$nb_{in}$ edge particles  from the inner edge region and
$nb_{out}$ edge particles from the outer edge region.

\begin{figure}
\centerline{\includegraphics[width=3.1in,height=3.0in]{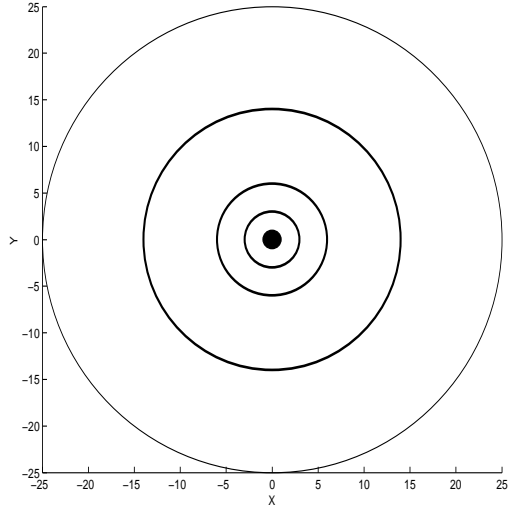}}
\label{subdom} \caption{domain subdivision}
\end{figure}

 In each iteration loop the $n_i$ particle of a sub
domain are evolved taking into account the presence of the all
$nb$ particles (for density, pressure forces etc. calculation).
Instead the $nb$ particles are evolved according to the kinematic
and dynamic properties they had as internal particles of their
original domain. At the end of the time step and before to start
the next one a check is performed to verify if some internal
particle has migrated into another domain. In this case the
particles data are exchanged.

\section{Results and data analysis}

We show the basic results of the simulations using one panel for
each case. In each panel we have 4 figures: the figure top-left
shows the number of particles versus time; the figure top-right
shows the radial Mach number of all the particles folded, in
$\varphi$, around the z axis; the figure bottom-left shows the
radial speed; the figure bottom-right shows the distribution of
the angular momentum of the particles, the scattering of the
values is just due to the conservation of the total angular
momentum.
\begin{table}
  \centering
  \textbf{Panel 1.} Simulation with f=0.0
    \label{rotazione 0pc}
  \begin{tabular}{|c|c|}
  \includegraphics[width=2.88in,height=2.04in]{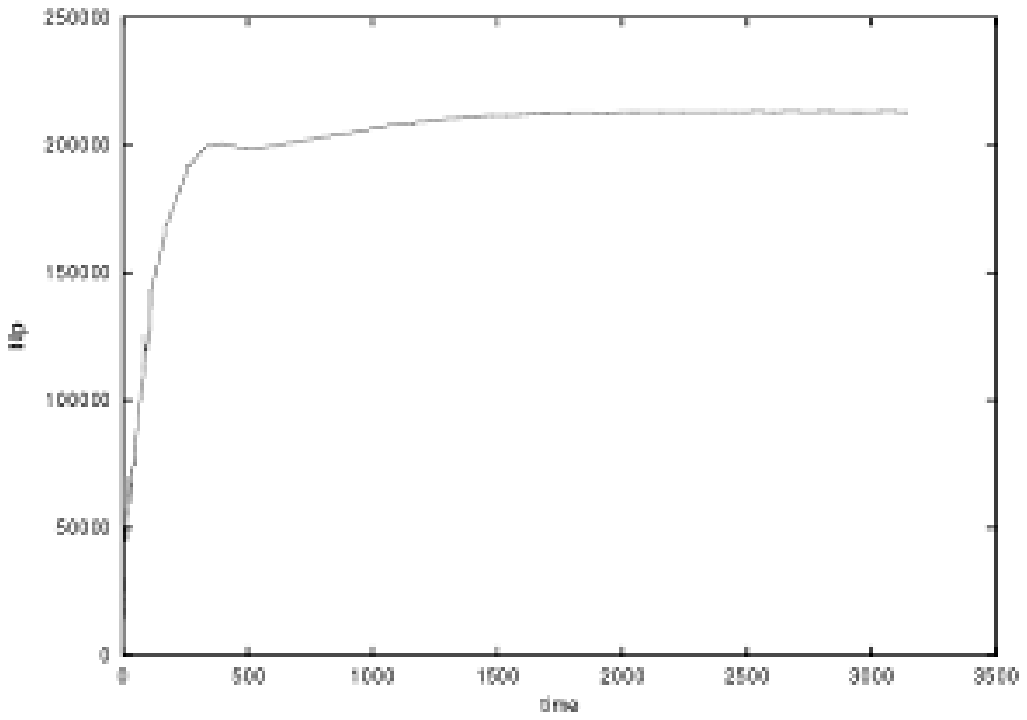} &  \\
    \textbf{Figure 1.1} Number of particles versus time & \\
    \includegraphics[width=2.88in,height=2.04in]{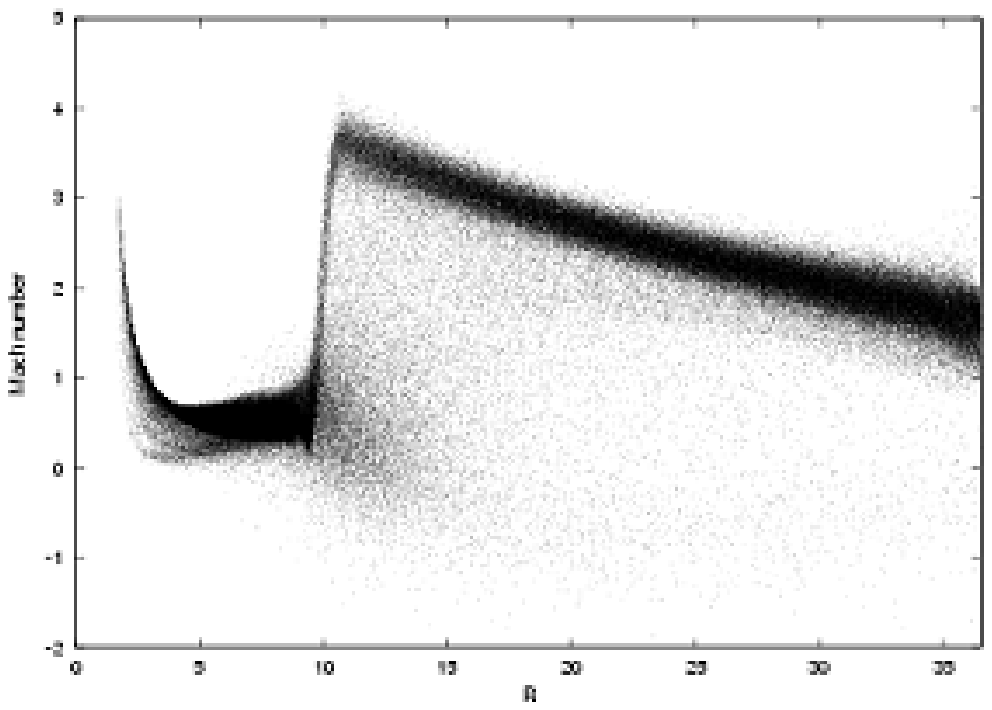}& \\
   \textbf{Figure 1.2} Radial Mach Number versus R\\
   \includegraphics[width=2.88in,height=2.04in]{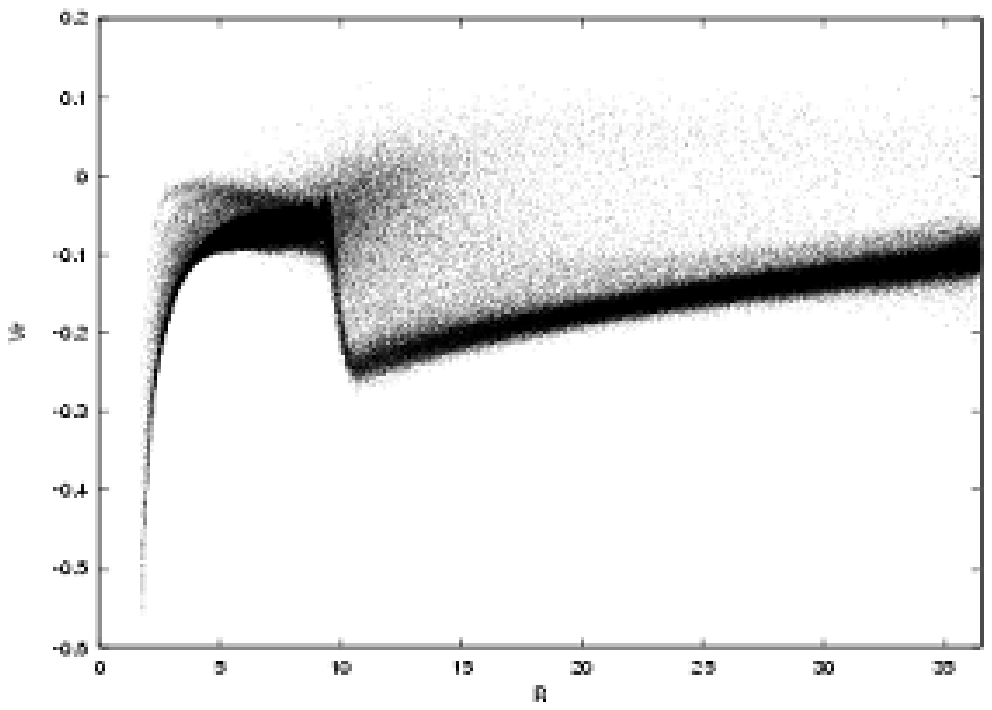} &  \\
    \textbf{Figure 1.3}  Radial speed versus R & \\
    \includegraphics[width=2.88in,height=2.04in]{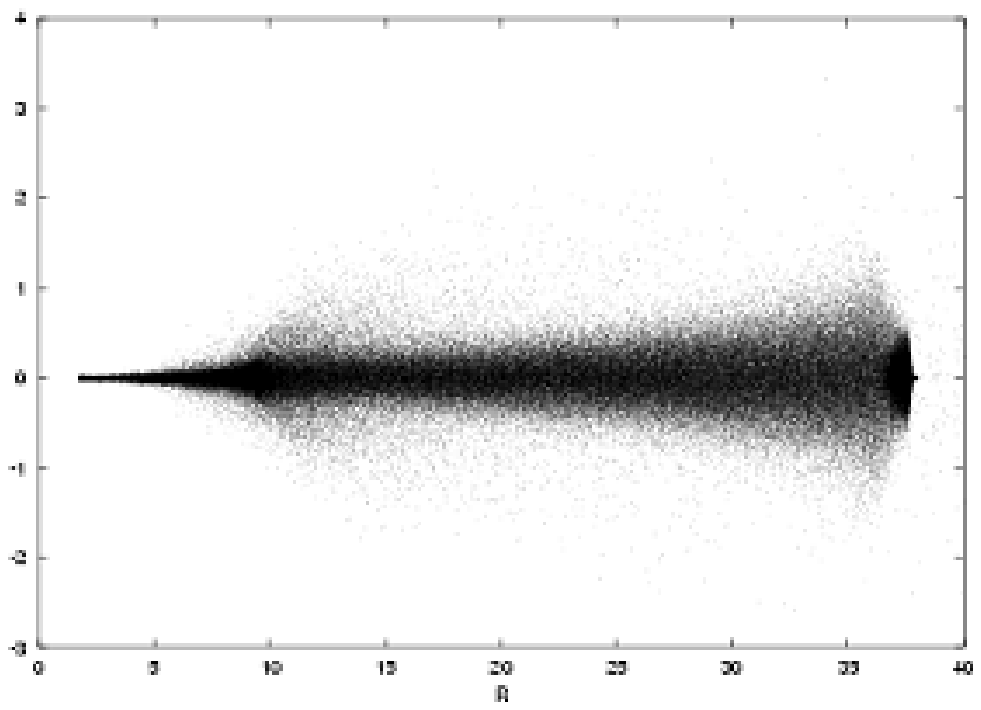}& \\
    \textbf{Figure 1.4} True angular momentum versus R & \\
\end{tabular}
\end{table}

As it is clear from Figure 1.1 the system reaches the equilibrium.
Figure 1.2, with the folded radial Mach number values, shows that
the shock is located at $R_{shock}=10.5$, a position close to the
theoretical one ($R_{shock}=6.3$). A not perfect agreement between
the theoretical shock position and the simulation result is not
surprising: a discrepancy has to be expected due to the vertical
motion of the gas in 3D geometry, while the theory assumes
vertical equilibrium (the Mach number profile is also very close
to the theoretical one). Figure 1.4 shows that also the mean value
of $\lambda$ is close to the real zero value.

We then simulated the same flow but increasing the amount of true
rotation. For a fractio $f=0.1$ the flow still reaches the
equilibrium state, but the shock is located somewhat outwards
compared to the theoretical position.

\begin{table}
  \centering
  \textbf{Panel 2.} Simulation with f=0.1
  \label{rotazione 10pc}
  \begin{tabular}{|c|}
  \includegraphics[width=2.88in,height=2.04in]{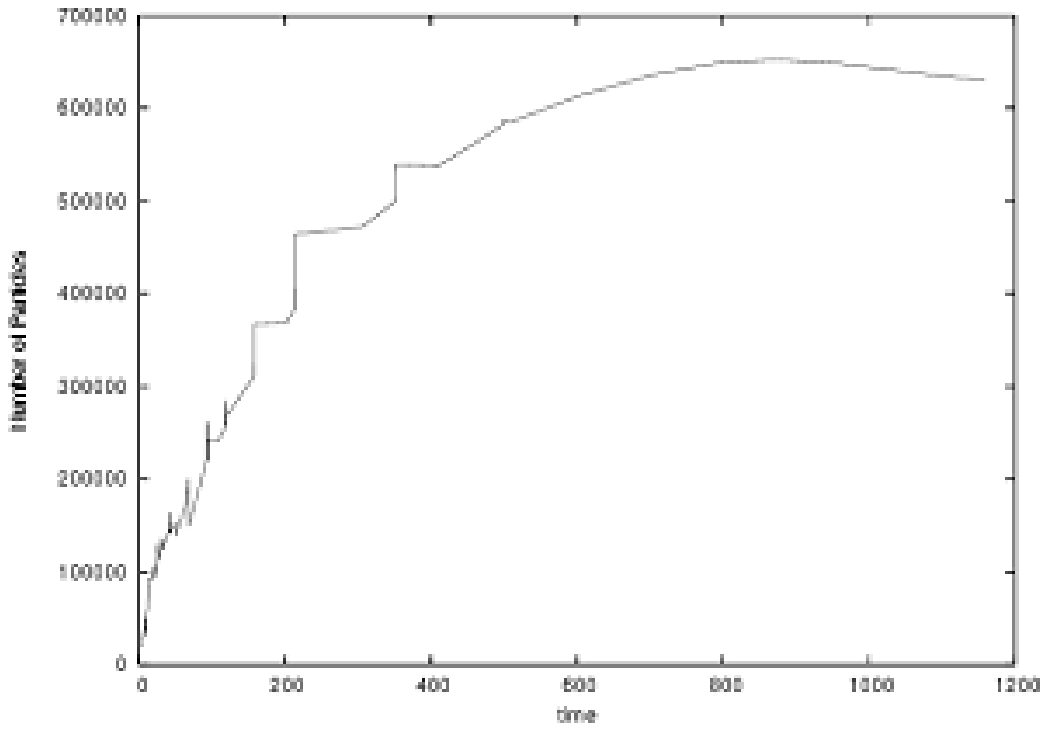}   \\
   \textbf{Figure 2.1} Number of particles versus time   \\
  \includegraphics[width=2.88in,height=2.04in]{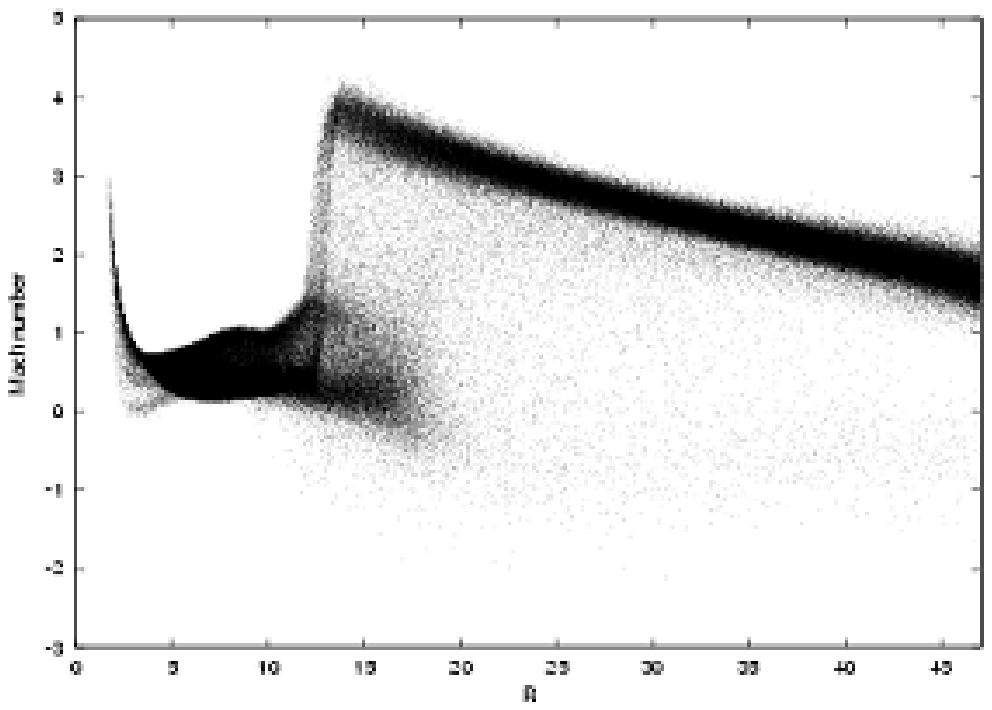} \\
  \textbf{Figure 2.2} Radial Mach Number versus R   \\
  \includegraphics[width=2.88in,height=2.04in]{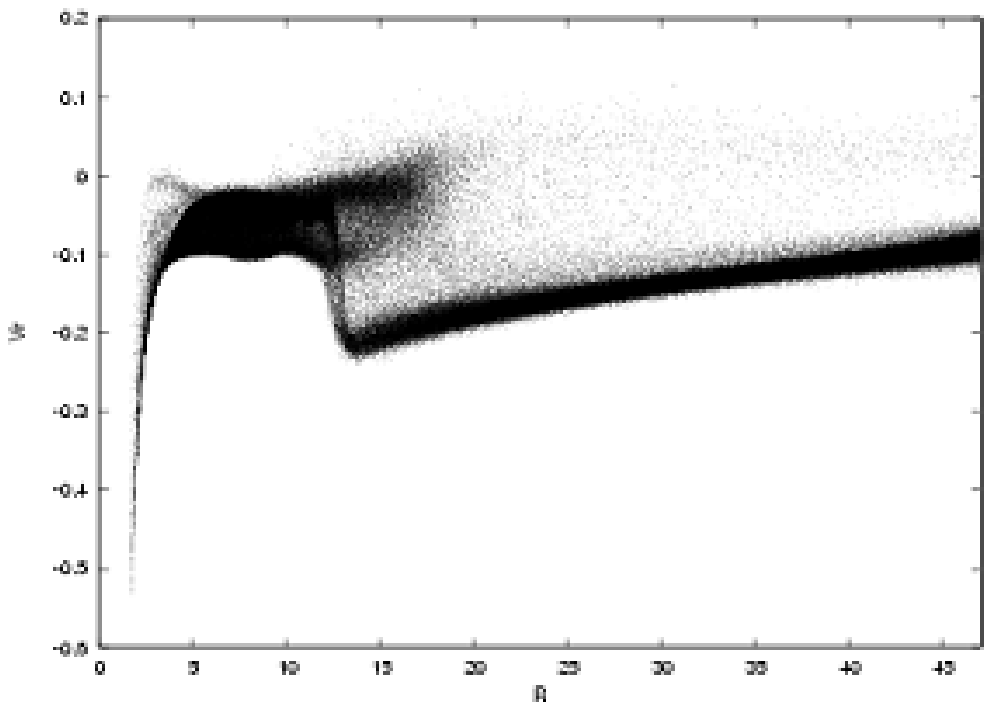}   \\
   \textbf{Figure 2.3}  Radial speed versus R  \\
  \includegraphics[width=2.88in,height=2.04in]{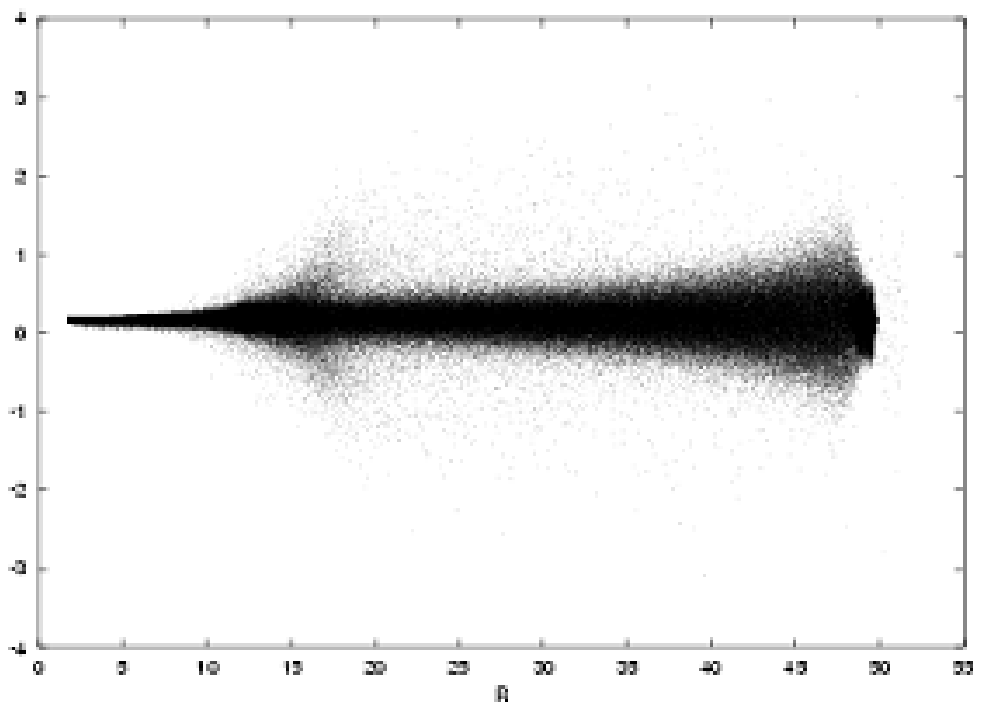}  \\
  \textbf{Figure 2.4} True angular momentum versus R  \\
\end{tabular}
\end{table}


 For  $f=0.2$ the results are shown in Panel 3. The
shock is steady, but at a larger radius.

For $f=0.4$, see Panel 4,  the system is still  in steady state
and the shock is further shifted outwards. It is clearly seen that
the number of particles versus time is increased.

For $f\geq 0.7$ (not shown) the shock position is unstable: it
travels steadily outwards. The number of particles is always
increasing. We stopped the simulation before the shock reached the
outer boundary of our integration domain.

In general the figures showing the radial speed versus $r$ point
out that, as the true rotation increases, the number of outflowing
particles increases. We inject with a fixed angular momentum, but
the numerical code viscosity, while conserving the total angular
momentum, produces an angular momentum redistribution and
therefore the particles with large angular momentum flow away.
\\

\pagebreak[4]

\begin{table}
  \centering
  \textbf{Panel 3.} Simulation with f=0.2
  \label{rotazione 20pc}
  \begin{tabular}{|c|c|}
  \includegraphics[width=2.88in,height=2.04in]{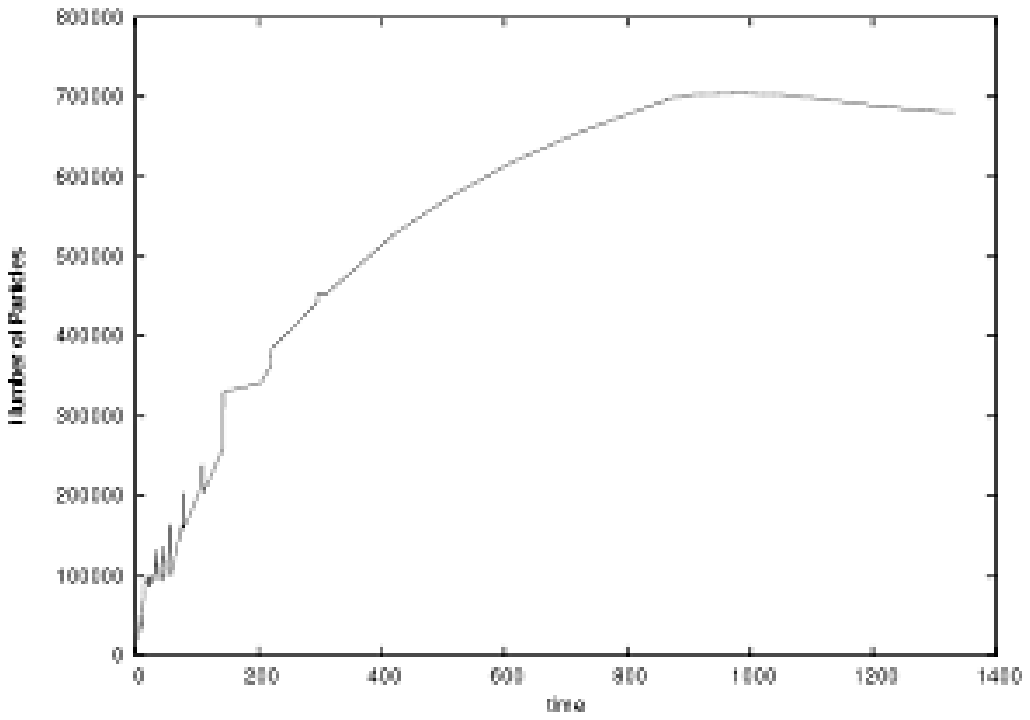} &  \\
    \textbf{Figure 3.1} Number of particles versus time & \\
 \includegraphics[width=2.88in,height=2.04in]{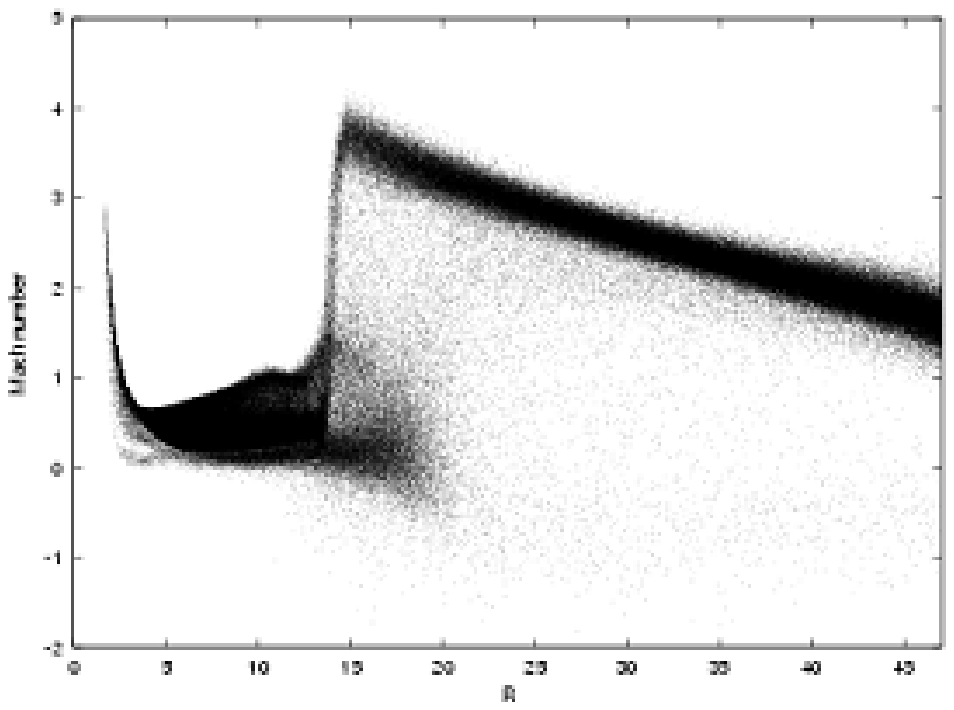} &\\
\textbf{Figure 3.2} Radial Mach Number versus R &\\
    \includegraphics[width=2.88in,height=2.04in]{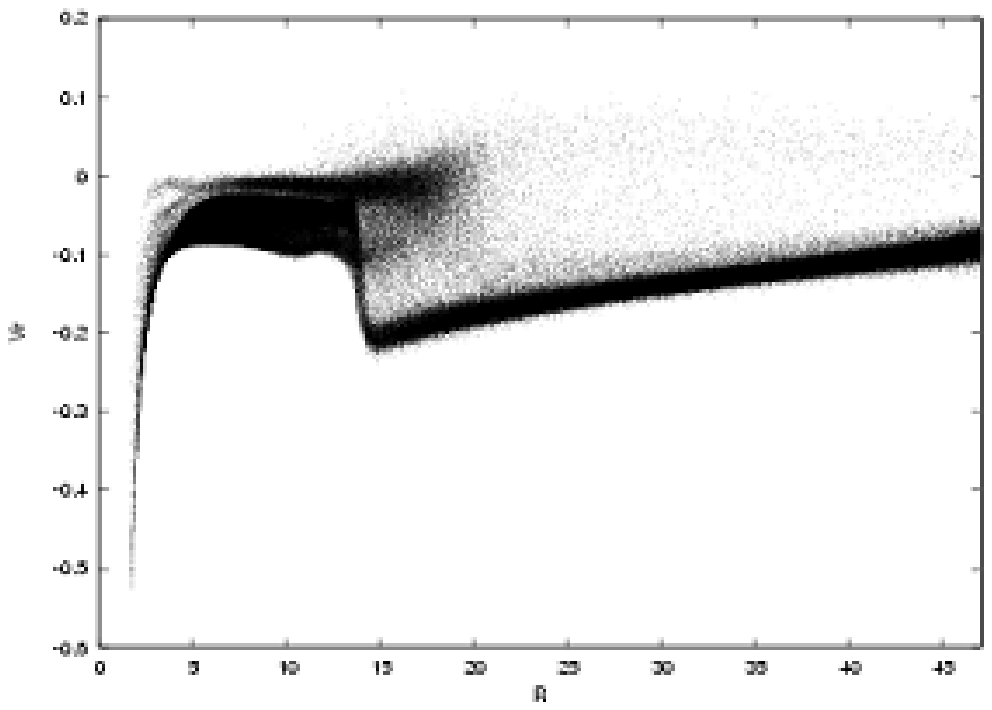} &  \\
    \textbf{Figure 3.3}  Radial speed versus R & \\
    \includegraphics[width=2.88in,height=2.04in]{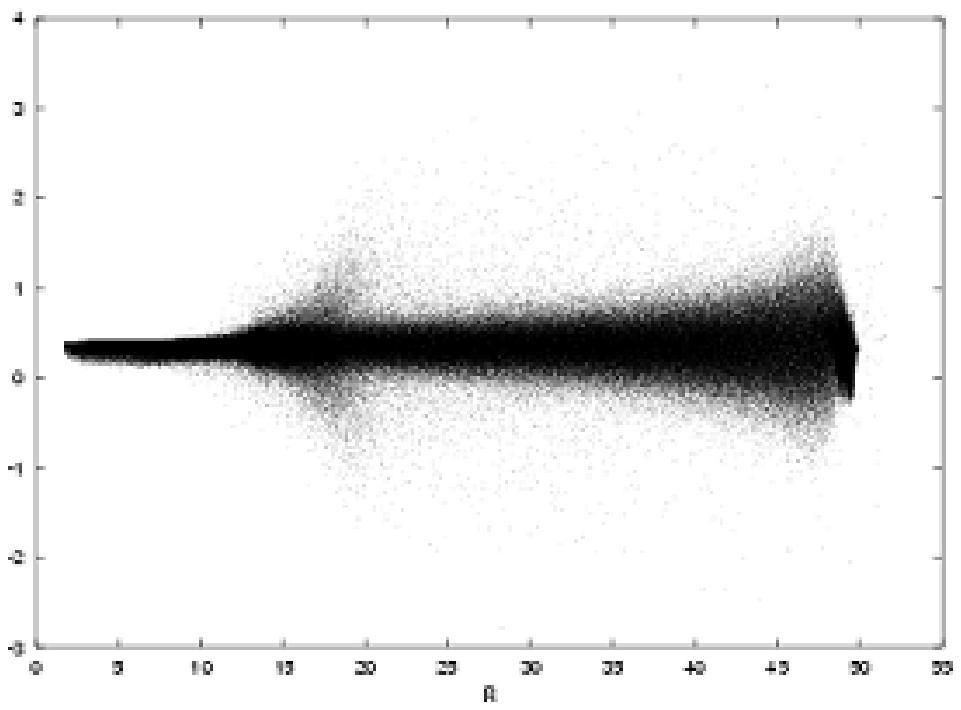} & \\
\textbf{Figure 3.4} True angular momentum versus R &\\
\end{tabular}
\end{table}

\begin{table}
  \centering
  \textbf{Panel 4.} Simulation with f=0.4
  \label{rotazione 40pc}
  \begin{tabular}{|c|c|}
  \includegraphics[width=2.88in,height=2.04in]{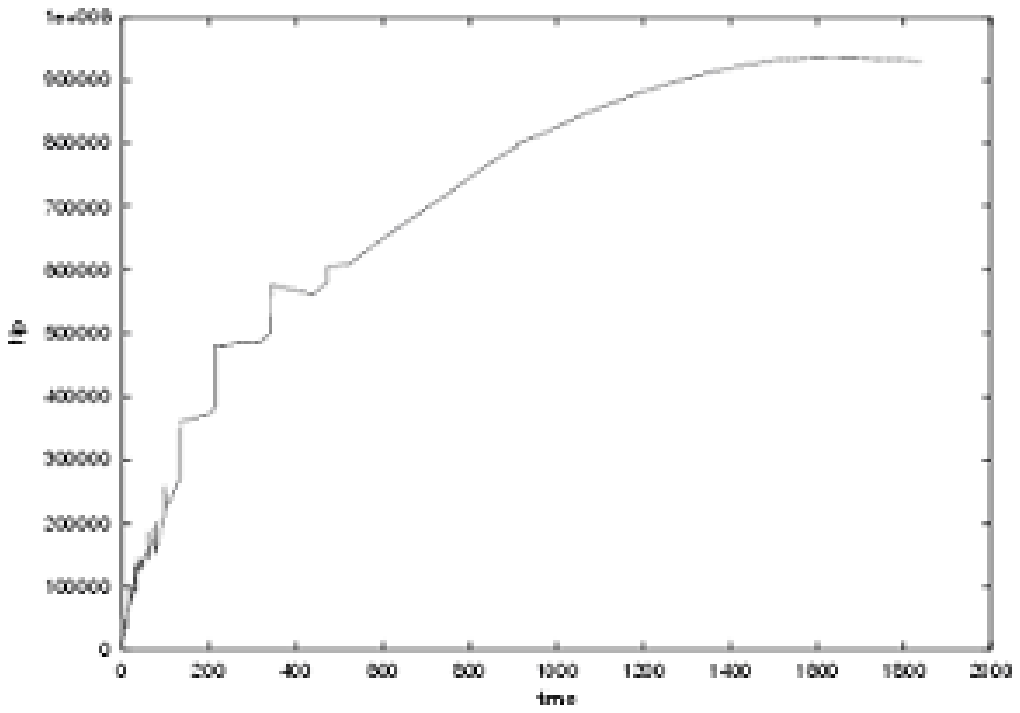}&   \\
   \textbf{Figure 4.1} Number of particles versus time & \\
 \includegraphics[width=2.88in,height=2.04in]{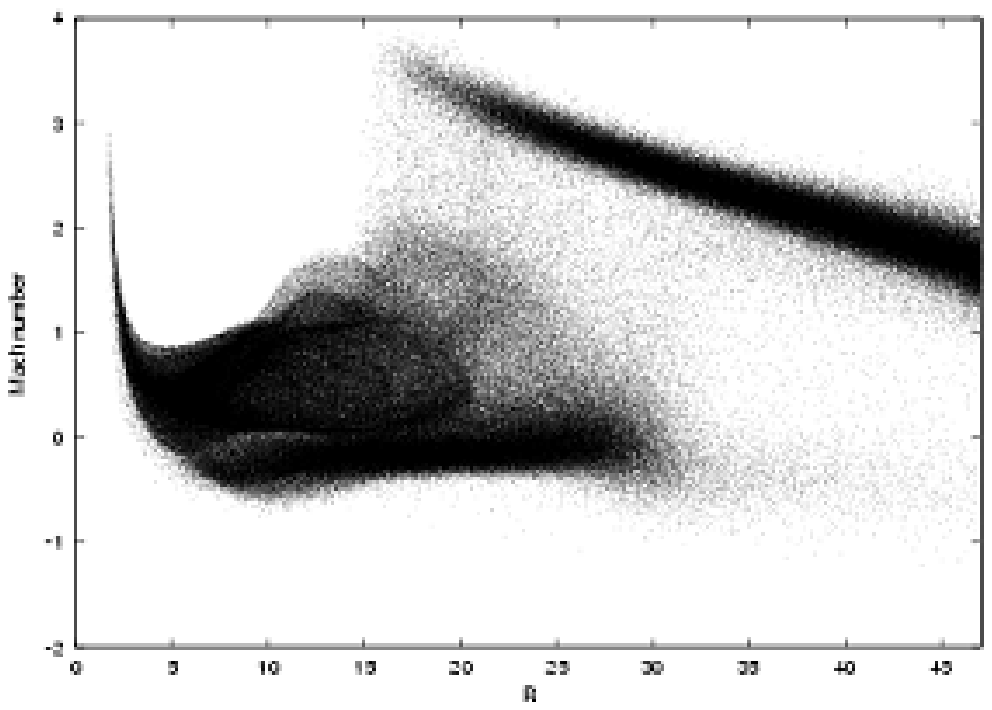} & \\
\textbf{Figure 4.2} Radial Mach Number versus R &\\
  \includegraphics[width=2.88in,height=2.04in]{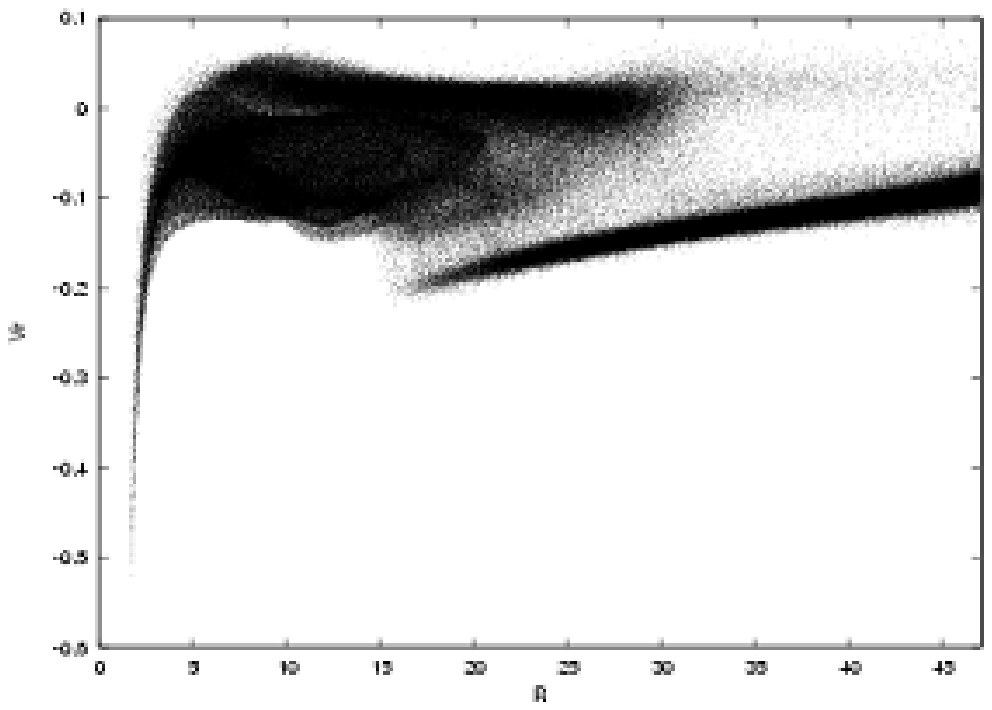} &  \\
  \textbf{Figure 4.3}  Radial speed versus R &\\
  \includegraphics[width=2.88in,height=2.04in]{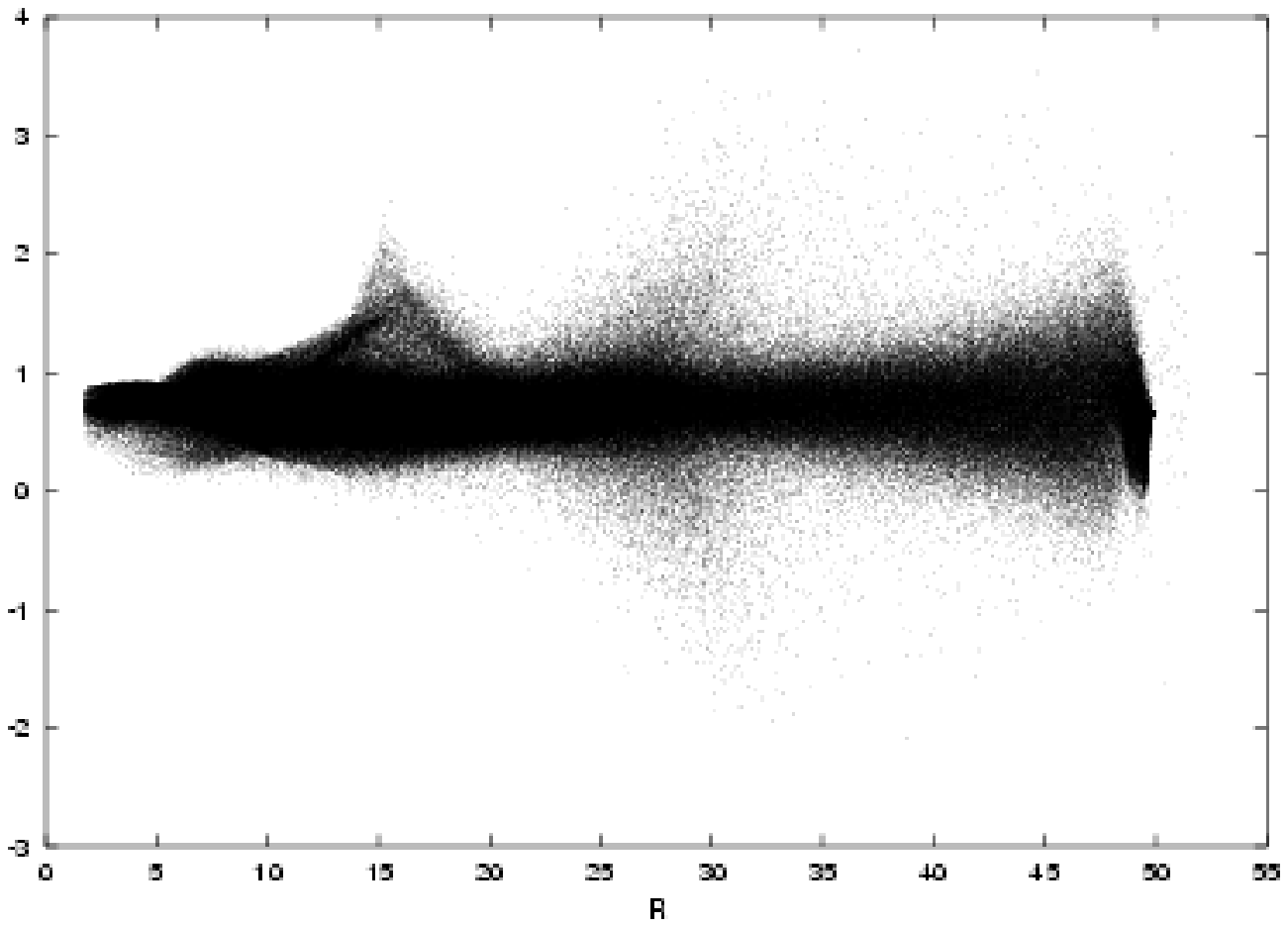} &\\
 \textbf{Figure 4.4} True angular momentum versus R &\\

\end{tabular}
\end{table}

\begin{table}
  \centering
  \textbf{Panel 5.} Simulation with f=0.4
  \label{rotazione 40pc}
  \begin{tabular}{|c|}
  \includegraphics[width=2.1in,height=2.5in,angle = 270.0]{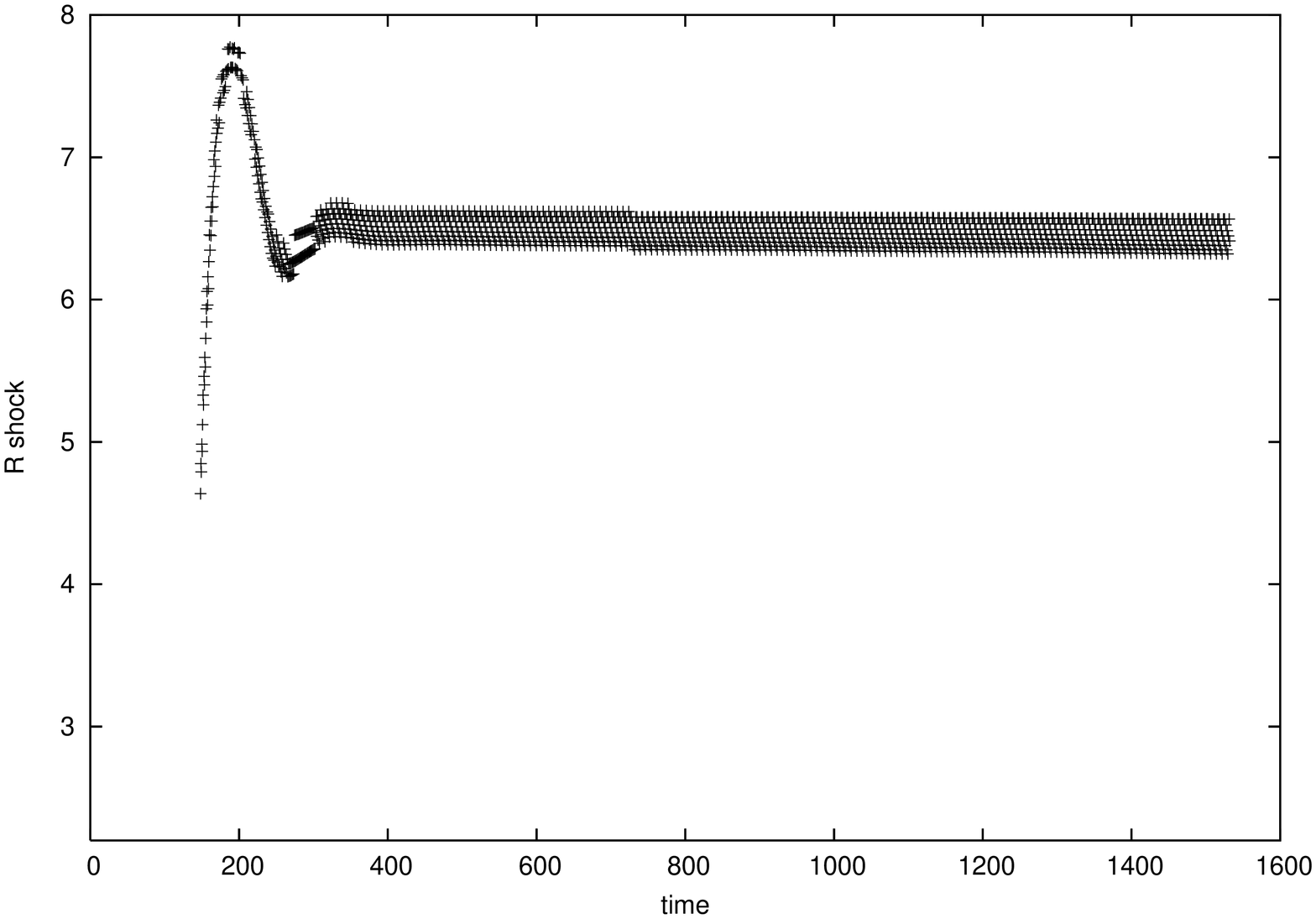} \\
   \textbf{Figure 5.1} Shock position versus time \\
\includegraphics[width=2.1in,height=2.5in,angle = 270.0]{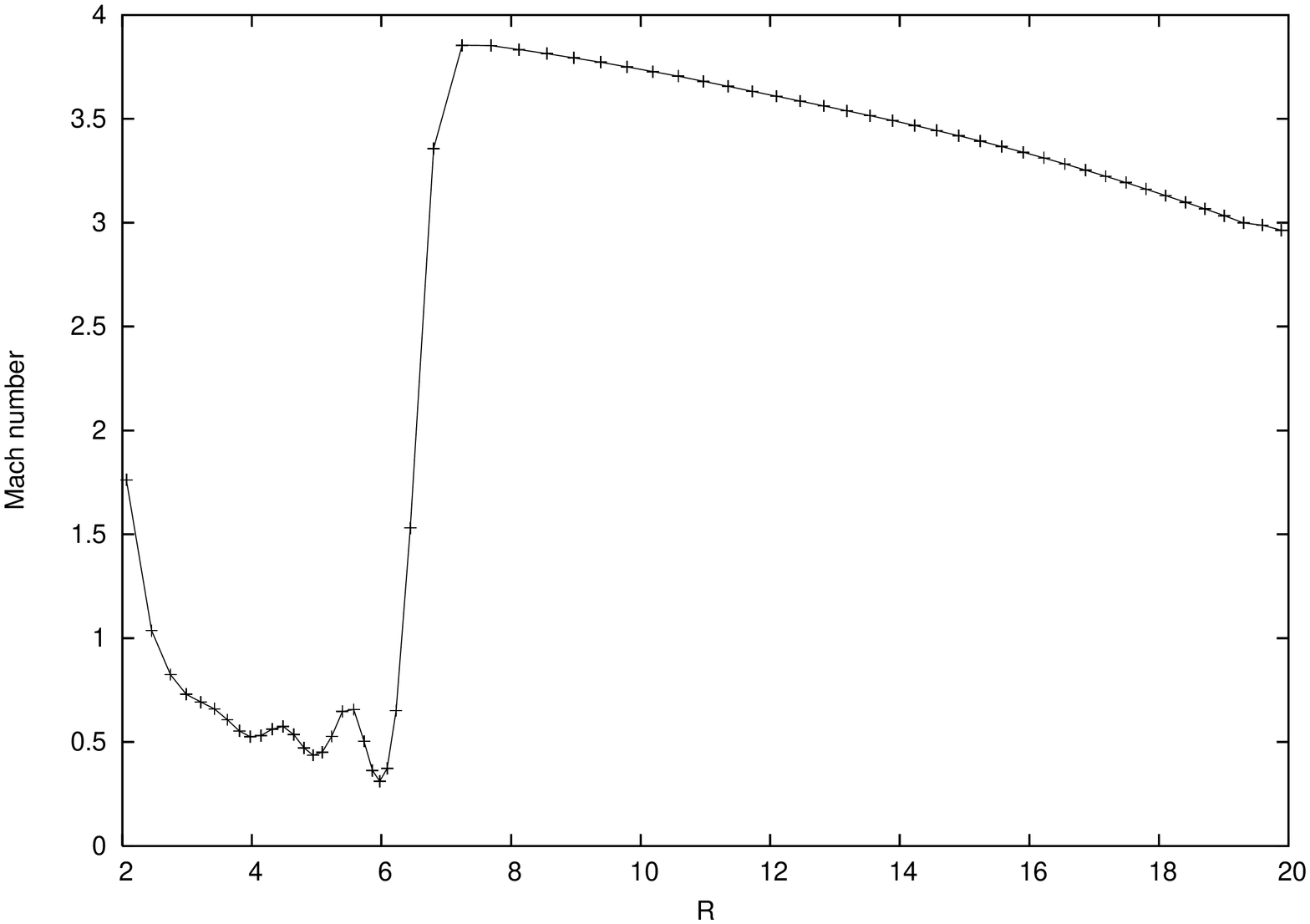} \\
\textbf{Figure 5.2} Radial Mach Number versus R\\
  \includegraphics[width=2.1in,height=2.5in,angle = 270.0]{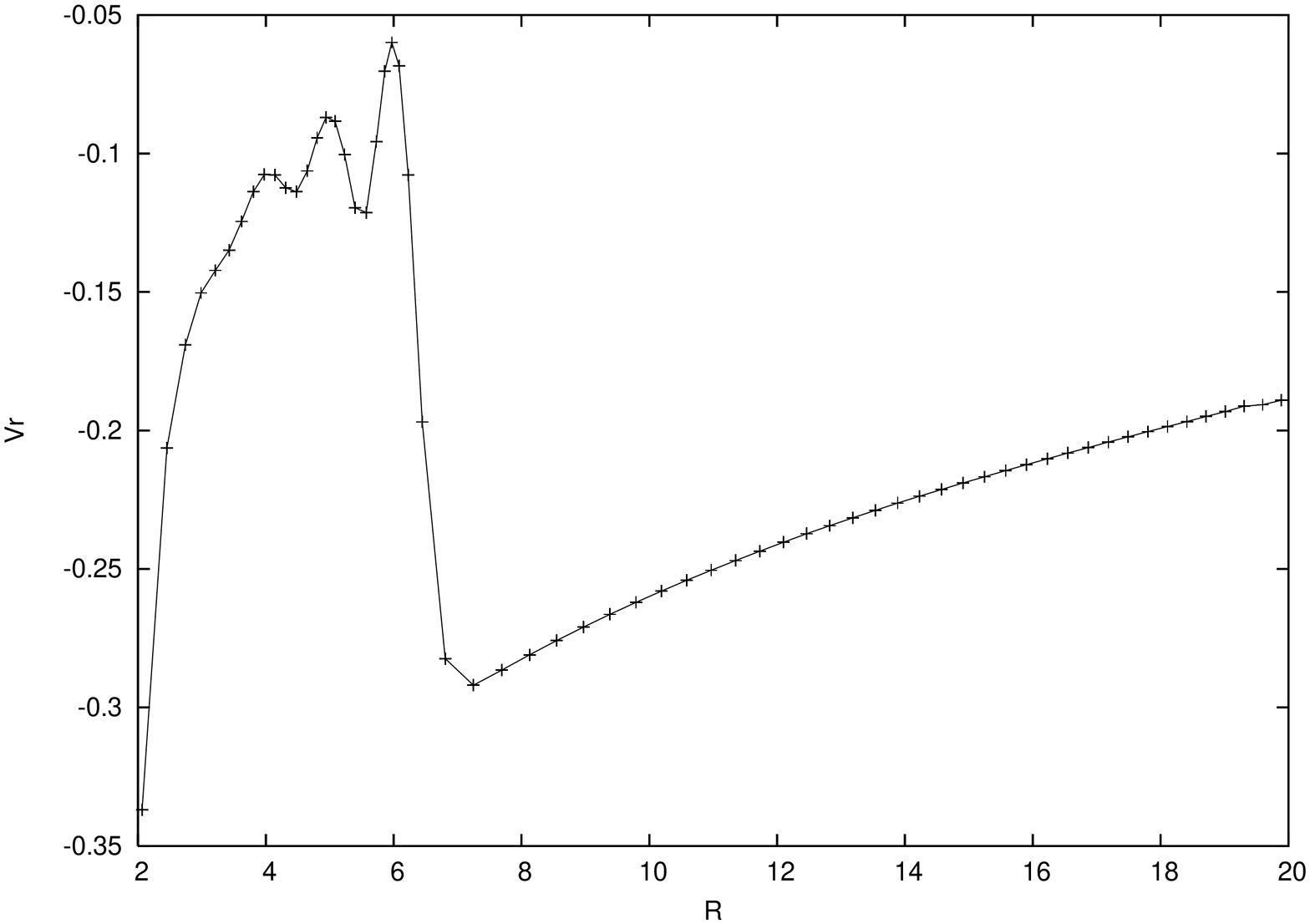} \\
  \textbf{Figure 5.3}  Radial speed versus R \\
 \includegraphics[width=2.1in,height=2.5in,angle = 270.0]{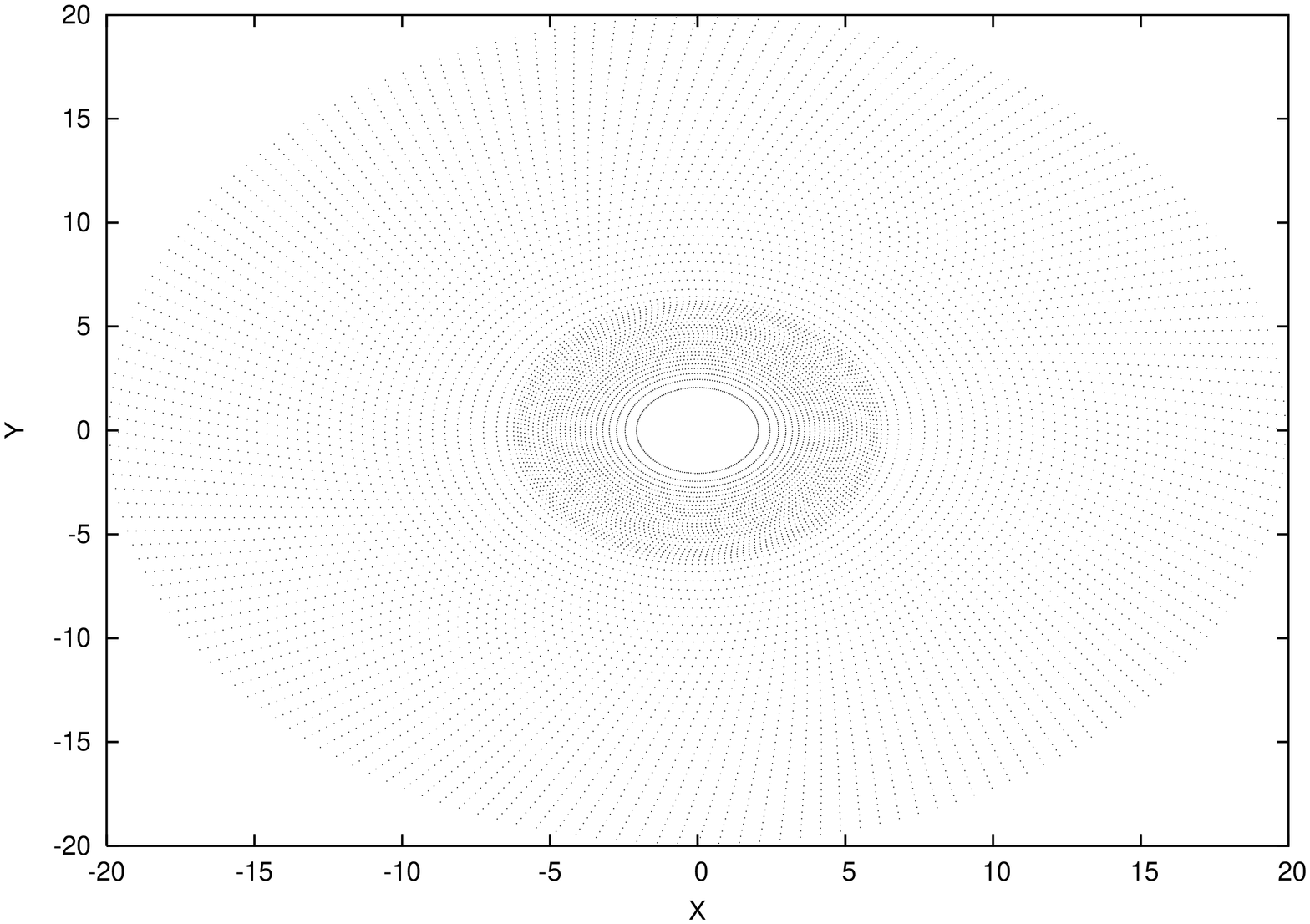}\\
 \textbf{Figure 5.4} Particle distribution in the XY plane\\
\end{tabular}
\end{table}
We made also 2D simulations, on the $Z=0$ plane. In this case the
agreement between the simulation results and the theory should be
perfect; the discrepancies are fully attributable to the code
properties. We checked that the 2D SPH simulations with different
f values show clearly the effect of the numerical shear viscosity.
Panel 5 shows the results obtained with $f=0.5$. The shock is
stable. In figure 5.1 the shock position versus time is plotted,
in figure 5.2 the radial Mach number (folded in $\varphi$) is
plotted versus R, in figure 5.3 the radial speed  (folded in
$\varphi$) is displayed and in in figure 5.4 the particle
distribution in the XY plane is presented.  In this case the
agreement between theory and simulation is quite good. Note that
the symmetry of the flow is so well conserved that the folded data
appear as one dimensional in R. The physical parameters are
$E=0.002$, $\lambda=1.8$. For gas with $\gamma=5/3$ the shock must
be $R_{shock}=6.78$, in good agreement with the simulation
results. The oscillations in Figure 5.1 are due to the algorithm
of the shock identification. The SPH numerical parameters are
$h=0.2$ and we use $\alpha=1$, $\beta=2$ for the artificial
viscosity coefficients. In the case with f=0.75, here not shown,
the shock is still stable, but the position is different from the
predicted one. In general, if we compare 2D and 3D simulations
with the same parameters and resolution, the 3D case is more
stable than the 2D one. We interpret this fact as due to the
impossibility of ejecting (along Z) high angular momentum gas
outside the domain in the 2D case, while in the 3D case the high
angular momentum particles are vertically ejected leading to a
stabilizing effect.

\section{Conclusions}

We showed that even in 3D time dependent simulations the standing
shocks predicted by Chakrabarti's analysis  are formed. In the
numerical simulations, for values of the parameters in the
theoretical stable zone, the shocks are stable. This fact confirms
the relevance of the shock phenomena in accretion flows around
black holes in galactic and extragalactic sources. Indeed the
standard keplerian disk models around galactic black hole
candidates suffer instabilities that make not completely clear the
physical scenario. In the case of the extragalactic sources like
AGN (active galactic nuclei), supposed to contain black holes,
while the keplerian disk can easily take account of the low energy
emission, the high luminosity at high energies, X and $\gamma$, is
still an open problem.

Due to the high CPU cost of 3D simulations the variability
properties of the shock flow exhibited in the reduced dimensional
simulations (2D in r-z axis symmetric configuration and 2D XY
plane configuration) will be the subject of a subsequent paper.

We showed also that this physical problem can be used as a test
for any fluid dynamics code since it is possible to derive the
analytical profile of the solution of the flow and compare it with
the simulated one for different degrees of real rotation of the
fluid. In this way it is possible to measure the numerical shear
viscosity of the code. We show that the 3D SPH code detects the
shock at a good level of accuracy.

\section{acknowledgements}
The simulations have been performed at the italian CINECA center for super-computation. We are particularly
grateful to Claudio Gheller for assistance in debugging the parallelized SPH code.


\newpage

\begin{table}
\centering \caption{} \label{A}
\begin{tabular}{|p{36pt}|p{36pt}|p{36pt}|p{36pt}|p{36pt}|p{36pt}|p{36pt}|p{36pt}|p{36pt}|}
\hline & r$_{i}$& v$_{ri }$ & a$_{ri}$ & $\lambda$ & H & f & R$_{shock}$&B \\
\hline I& 38 & -0.076 & 0.069&  1.625 &16 &0.0 &5.76 & 0.0045 \\
\hline II& 50 & -0.072 & 0.058& 1.65  & 20 &0.1 & 6.32& 0.003 \\
\hline III& 50 & -0.072 & 0.058& 1.65  & 20 &0.2 & 6.32& 0.003 \\
\hline IV& 50 & -0.072 & 0.058& 1.65  & 20 &0.4 & 6.32& 0.003 \\
\hline
\end{tabular}
\end{table}

\label{lastpage}

\end{document}